\documentclass[%
reprint,
amsmath,amssymb,
aps,
]{revtex4-2}
\usepackage[colorlinks=true,allcolors=blue]{hyperref}
\usepackage{hyperxmp}
\usepackage{bm}
\usepackage{float}
\usepackage{booktabs,tabularx,array}
\usepackage{graphicx,color}
\usepackage[dvipsnames]{xcolor}

\usepackage{enumerate}
\usepackage{harpoon}
\usepackage{amsmath,amssymb}
\usepackage{gensymb}
\usepackage{accents}
\usepackage[e]{esvect}
\usepackage[type={CC},modifier={by},version={4.0}]{doclicense}
\usepackage{footmisc}

\definecolor{ctcolor}{rgb}{0.8, 0.0, 0.2}

\usepackage{dcolumn}
\usepackage{bm}
\usepackage{dsfont}

\usepackage{graphicx}
\usepackage{dcolumn}
\usepackage{bm,amsmath}
\usepackage{xcolor}
\usepackage{hyperref}

\begin{document}
\title{Exceptional Topological Signatures of Non-Hermitian Photonic Hopf-Link Braids}

\author{Samit Kumar Gupta}
\email{samitg@iiserbpr.ac.in\,, samit.kumar.gupta@gmail.com}
\affiliation{Department of Physical Sciences, Indian Institute of Science Education and Research Berhampur, Odisha 760003, India}

\begin{abstract}
Non-Hermitian physics endows the non-Abelian systems with exceptional topology characterized by non-commutative braid patterns. Interplay of distinct competing sources of non-Hermiticity may induce novel topological effects. Here, we provide a generalized Hatano-Nelson model with higher-order nonreciprocal hoppings, non-Abelian gauge fields, and staggered gain-loss processes showing the exceptional topological structure of the Hopf-link braids that undergoes a EP-mediated topological phase transition. We demonstrate that mixing multiple nonreciprocal channels drives the system into highly intricate, nested complex energy Hopf-link braids and expands the topological landscape up to higher-order braiding sectors. Furthermore, utilizing biorthogonal eigenvector tracking, we map the structural evolution of the exceptional phase boundaries via the maximum Petermann factor in the phase angle parameter planes. We show that the gain-loss non-Hermiticity drives a topological crossover where the extended exceptional contours constrict into isolated regimes. The demonstration of EP-mediated topological phase transition of Hopf-link braids and the associated rich exceptional phase portraits may offer new physical insights with promising applications in robust, fault-tolerant communication channels and quantum computing platforms.

\end{abstract}
\maketitle
\emph{Introduction.}--- Non-Hermitian physics has emerged to shed new light in optical, photonic, condensed matter systems \cite{el2018non,feng2017non,ozdemir2019parity,konotop2016nonlinear,gupta2020parity} mainly due to the non-Hermitian topology and exceptional points (EPs) \cite{ashida2020non,li2023exceptional,ghatak2019new,torres2019perspective,miri2019exceptional} revealing intriguing physical effects and phenomena in the theoretical \cite{bergholtz2021exceptional,ding2022non,leykam2017edge} and experimental \cite{ruter2010observation,peng2016chiral,chen2020revealing} realms, most commonly including topological winding around non-Hermitian singularities \cite{zhong2018winding,midya2018non, zhang2020correspondence} and topological signatures of non-Hermitian skin effects (NHSE) \cite{martinez2018non,yao2018edge,kunst2018biorthogonal,jin2019bulk,okuma2020topological,zhang2022review,zhu2020photonic,okuma2023non,lin2023topological, li2020critical,qin2023universal,wang2023scaling, yuce2021nonlinear,shen2022non, zhang2021observation, zhang2022universal}. On the other hand, non-Hermitian topology of exceptional points is inherently related to the non-Abelian braiding of complex energy \cite{wang2021topological,noh2020braiding,guo2023exceptional,zhang2022non,patil2022measuring,yang2024non,guria2024resolving, midya2023gain} and knotted topological structures \cite{hu2021knots,cao2024observation, jiang2026generating}, including the recent realization of the braiding of non-Hermitian laser modes in integrated photonic chips \cite{mao2026laser, konig2026braids}.

Recently, non-Abelian physics has revealed promising new effects and phenomena, including non-Abelian anyons and energy braiding \cite{nayak2008non, stern2010non, zhang2022non}. The implementation of non-Abelian gauge fields in photonic systems has established synthetic gauge field as a practical tool to achieve robust topological transport \cite{chen2019non, zhang2022non, yang2024non}. To realize synthetic SU(2) gauge fields, various approaches have been employed including anisotropic metamaterials \cite{chen2019non, yang2019synthesis}, dynamically modulated photonic synthetic-frequency lattices utilizing photon polarization as internal spin and modulation-induced couplings as matrix-valued SU(2) hopping phases \cite{cheng2025non}, integrated photonic chips based on tunable waveguide networks \cite{zhang2022non, sun2025reconfigurable, jiang2026photonic,neef2023three}, and free-space optics \cite{wong2025synthetic}. In these platforms, an internal degree of freedom of two-modes SU(2) is achievable via polarization, dual spatial modes, or two-resonator \textit{photonic molecules} in waveguide arrays \cite{zhang2022non}. 

Synthetic gauge photonic lattices based on non-Abelian gauge fields may be fertile grounds for exploring the non-Abelian dynamics of complex energy bands \cite{pang2024synthetic}. In such cases, considering matrix-valued non-Abelian couplings may prove to be advantageous in inducing a higher-order braiding degree \cite{chen2026implementing}. Naturally, simultaneous consideration of physically different competing sources of non-Hermiticity and gain-loss-induced exceptional points may play an important role in inducing a rich tapestry of complex topological phenomena. However, a complete picture is still lacking despite some developments, including the depiction of gain-loss EP-mediated topological phase transition of a necklace of Hopf-links with NNN interactions \cite{gupta2025non} and experimental realization of Hopf-link braiding with nearest neighbor (NN) interaction in electric circuits \cite{chen2026implementing}. In addition, we note that Hopf phases of matter in two-band topological systems are usually delicate and unstable in generic non-Hermitian settings \cite{pak2024pt,yang2019non,zhang2020bulk, kim2023realization, wojcik2020homotopy, nakamura2025non}. Homotopy classification of Hopf bundle is affected by non-Hermitian degrees of freedom. However, $\mathcal{PT}$ symmetry has been found to stabilize the Hopf topological invariants despite the presence of non-Hermiticity \cite{pak2024pt}. 

In this work, we consider a 1D tight-binding (TB) lattice based on the Hatano-Nelson model under periodic boundary condition (PBC) with imbalanced hopping amplitudes and non-Abelian gauge phases up to third-order (next-to-next nearest neighbor, NNNN), and staggered onsite gain-loss potentials. Mixing the higher-order nonreciprocal interactions unlocks highly intricate nested complex energy Hopf-links expanding the topological phase diagram with higher-order braiding sectors. Utilizing bi-orthogonal eigenvector tracking, we reveal the structural evolution of the exceptional phase boundaries via the maximum Petermann factor divergence in the non-Abelian phase parameter planes. We demonstrate that increasing gain-loss non-Hermiticity drives a topological crossover in which the exceptional contours are contracted into isolated EP regimes, reducing the Petermann factor to small background values. 


\emph{Theoretical model.}--- The real-space tight-binding Hamiltonian of the system based on the Hatano-Nelson model with U(1) Abelian phases can be written as:
\begin{figure}
    \centering
    \includegraphics[width=0.98\linewidth]{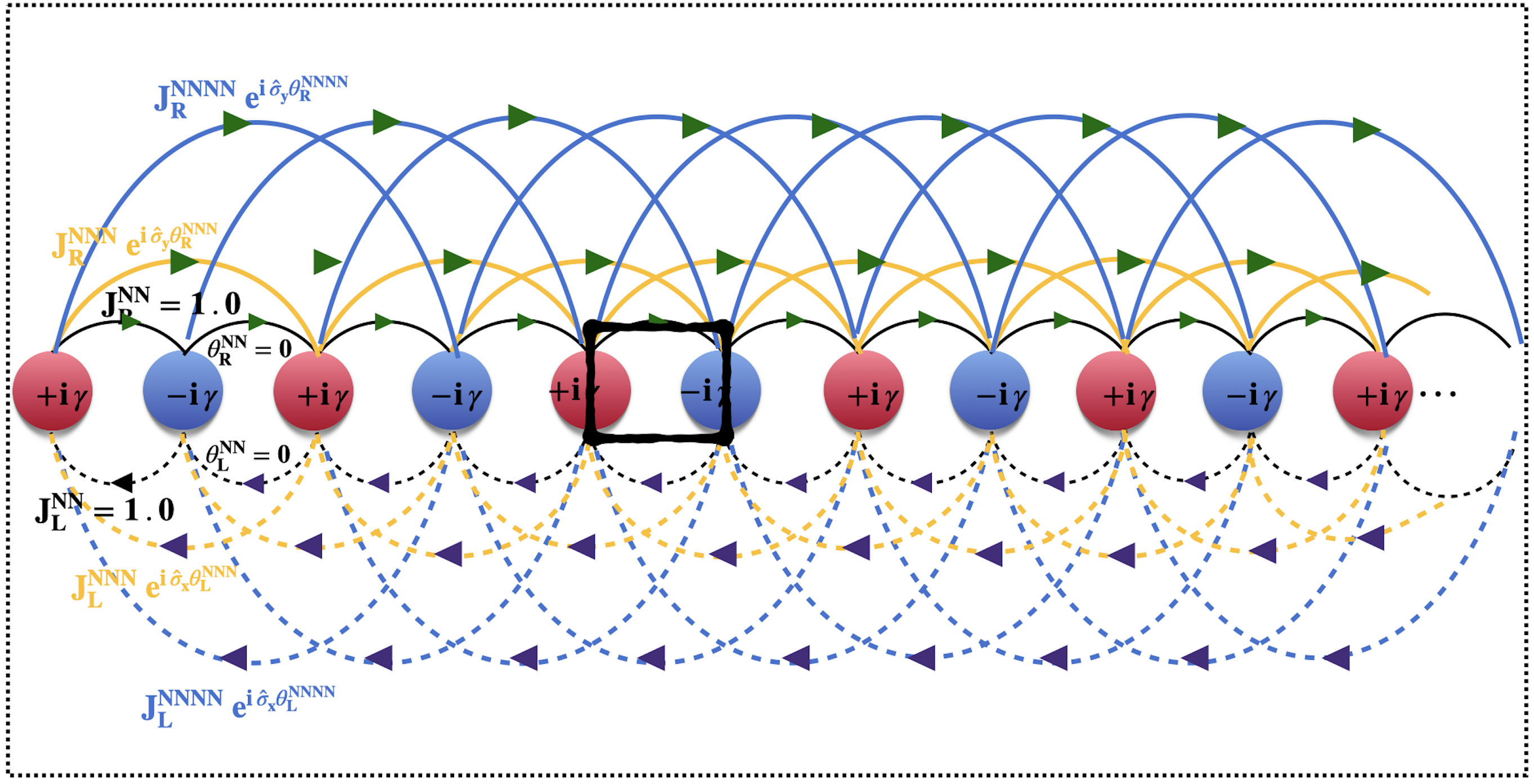}
    \caption{Schematic diagram of the system based on the 1D Hatano-Nelson model. The solid black lines represent right-ward NN hoppings whereas dashed black lines represent left-ward NN hoppings. The solid yellow lines represent right-ward NNN hoppings whereas dashed yellow lines represent left-ward NNN hoppings. The solid blue lines represent right-ward NN hoppings whereas dashed blue lines represent left-ward NN hoppings. The red and blue disks represent gain and loss sites respectively with a unit cell represented by a rectangular box at the center.}
    \label{fig1}
    \end{figure}
    
\begin{equation}
\hat{H}_{R}^{A}
=\sum_{m}{(J_{L}^{NN}\; {\hat{c}^{\dagger}_{m}}e^{i\theta_{L}^{NN}\,} \,{\hat{c}_{m+1}}+J_{R}^{NN} \;{\hat{c}^{\dagger}_{m+1}} e^{i\theta_{R}^{NN}\,} \,{\hat{c}_{m}})}, 
\end{equation}
where, $\theta_{L}^{NN}$, $\theta_{R}^{NN}$ are scalar Abelian phases. However, non-Abelian SU(2) phases induce intrinsic two-band spin structures such as $\hat\theta^{NN}_{L,R}=\theta^{NN}_{L,R}\;\hat\sigma_{y,x}$ and $\hat\theta^{NNN}_{L,R}=\theta^{NNN}_{L,R}\;\hat\sigma_{y,x}$ in Eq. \ref{eq:na} and Eq. \ref{eq:na_r_NNNN} , respectively. Including the non-Abelian SU(2) phases in the left-ward or right-ward nonreciprocal NN hoppings, the TB real-space lattice Hamiltonian becomes \cite{pang2024synthetic}: 
\begin{equation}{\label{eq:na}}
    H^{NA}_{R}=\sum_{m}{(J_{L}^{NN}\; \hat{c}^{\dagger}_{m}}e^{i\theta_{L}^{NN}\,\hat{\sigma}_{y}} \,{\hat{c}_{m+1}}+ J_{R}^{NN} \;\hat{c}^{\dagger}_{m+1} e^{i\theta_{R}^{NN}\,\hat{\sigma}_{x}} \,\hat{c}_{m}).
\end{equation}
In general, the corresponding real-space lattice Hamiltonian of the system with NN, NNN, and NNNN hoppings, non-Abelian SU (2) phases, and onsite gain-loss profiles can be written in the following second-quantized form: 
\begin{widetext}
\begin{align}
\hat{H}^{NNNN}_R&=\sum_{m}{(J_{L}^{NN}\; \hat{c}^{\dagger}_{m}}e^{i\theta_{L}^{NN}\hat{\sigma}_{y}} \;\hat{c}_{m+1}+J_{R}^{NN} \;\hat{c}^{\dagger}_{m+1} e^{i\theta_{R}^{NN}\hat{\sigma}_{x}} \;\hat{c}_{m})+\sum_{m}(J_{L}^{NNN} \; \hat{c}^{\dagger}_{m}e^{i\theta_{L}^{NNN}\hat{\sigma}_{y}}\; \hat{c}_{m+2}
+J_{R}^{NNN} \;\hat{c}^{\dagger}_{m+2} e^{i\theta_{R}^{NNN}\hat{\sigma}_{x}}\;\hat{c}_{m})\nonumber \\        
& \quad +\sum_{m}(J_{L}^{NNNN} \; \hat{c}^{\dagger}_{m}e^{i\theta_{L}^{NNNN}\hat{\sigma}_{y}}\; \hat{c}_{m+3}
+J_{R}^{NNNN} \;\hat{c}^{\dagger}_{m+3} e^{i\theta_{R}^{NNNN}\hat{\sigma}_{x}}\;\hat{c}_{m})+\sum_{m}{i \;\gamma(-1)^{m} \; \;\hat{c}^{\dagger}_{m} \hat{c}_{m}}
\label{eq:na_r_NNNN}
\end{align}
\end{widetext}

Here, the first term under summation on the right-hand side of Eq. \ref{eq:na_r_NNNN} refers to the interaction term due to NN hoppings, the second term under summation refers to the interaction term due to NNN hoppings, the third term under summation refers to the NNNN hoppings, and the last one is the staggered gain/loss term.
Its underlying physics can be captured by the following $2\times 2$ k-space effective Hamiltonian with SU(2) non-Abelian gauge fields under periodic boundary condition (PBC), when, for the sake of simplicity, we assume $\theta_{L}^{NN}= \theta_{R}^{NN}=0$ and $J_{L}^{NN}=J_{R}^{NN}=1.0$. The non-Abelian gauge fields in the NN and NNN hoppings in Eq. \ref{eq:na} and \ref{eq:na_r_NNNN} induce non-Hermiticity due to directional coupling. The onsite gain-loss distribution basically introduces an alternately gain-loss staggered imaginary gauge potential across the lattice: $V=i\,\gamma\, \times\;\textbf{diag}(1,\,-1,\,1\,,-1\,...(-1)^N).$ In general, simultaneous consideration of NN, NNN, and NNNN hoppings along with onsite gain-loss potentials induces increasing complexity in the system for which calculation of EP and related non-Hermitian topological dynamics becomes analytically intractable. Such a generic non-Hermitian photonic environment may lead to exotic complex wave phenomena previously unexplored. Long-range hoppings can, in general, enrich the spectral and localization properties of a system. The effective Hamiltonian turns out to be \cite{gupta2025non}:
\begin{widetext}
\begin{align}\label{eq:model_eq_eff_NNNN}
H^{NNNN}_{eff}(k)&=(A_{1}+A_{2}+A_3) \; \hat{\sigma}_{0}+ i \,\hat{\sigma}_{y}\,(J_{L}^{NNN} \; sin\theta_{L}^{NNN}\; e^{2ik}+J_{L}^{NNNN} \; sin\theta_{L}^{NNNN}\; e^{3ik}) \nonumber \\
& +i \, \hat{\sigma}_{x} \,(J_{R}^{NNN} \;sin\theta_{R}^{NNN} \; e^{-2ik}+J_{R}^{NNNN} \;sin\theta_{R}^{NNNN} \; e^{-3ik})+i\, \gamma\hat{\sigma}_{z},
\end{align}
\end{widetext}
where, $A_{1}=(J_{L}^{NN}e^{ik}\;cos\theta_{L}^{NN}+J_{R}^{NN}e^{-ik}\;cos\theta_{L}^{NN})$,
$A_{2}=(J_{L}^{NNN} \; cos\theta_{L}^{NNN}\; e^{2ik}+J_{R}^{NNN}$, 
$A_{3}=(J_{L}^{NNNN} \; cos\theta_{L}^{NNNN}\; e^{3ik}+J_{R}^{NNNN}cos\theta_{R}^{NNNN}\; e^{-3ik})$, $\hat{c}_{m}^{\dagger}$/ $\hat{c}_{m}$ are the creation/annihilation operators, respectively. Here, $\hat{\sigma}_{x}=[0,\,-1;\;1,\,0], \hat{\sigma}_{y}=[0,\,-i;\;i,\,0], \hat{\sigma}_{z}=[1,\,0;\;0,\,-1]$ are Pauli matrices and $\hat{\sigma}_{0}=\hat{I}=[1,\,0;\;0,\,1],$ $J_{L}^{NN}$ and $J_{R}^{NN}$ are left- and right-ward NN hoppings, $J_{L}^{NNN}$ and $J_{R}^{NNN}$ are left- and right-ward NNN hoppings, $J_{L}^{NNNN}$ and $J_{R}^{NNNN}$ are left- and right-ward NNNN hoppings, $\theta_{L}^{NN}$ and $\theta_{R}^{NN}$ are left- and right-ward NN non-Abelian SU (2) gauge phases, $\theta_{L}^{NNN}$, and $\theta_{R}^{NNN}$ are left- and right-ward NNN non-Abelian SU (2)phases, and $\theta_{L}^{NNNN}$, and $\theta_{R}^{NNNN}$ are left- and right-ward NNNN non-Abelian phases, where N represents the total number of sites, and $\gamma$ refers to the gain-loss distributions on site. 
In this case, however, the nonreciprocity in the hopping amplitudes and non-Abelian phases in the NN and NNN interaction can be relaxed to simplify the model in analytically accessing EPs. In other words,  $\theta_{L}^{NN}= \theta_{R}^{NN}=0$,  $J_{L}^{NN}=J_{R}^{NN}=1.0$, $J_{L}^{NNN}=J_{R}^{NNN}$, $\theta_{L}^{NNN}= \theta_{R}^{NNN}=0$. With this simplification, its eigenvalues are found to be as follows:   

\begin{align}  \label{eq:model_eq_eig_NNNN2}
E_{\pm}&=(A_{1}+A_{2}+A_{3})) \, \pm \, i \sqrt{(X_1^2+X_2^2+\gamma^2)},  
\end{align}  
\begin{figure*}
\centering
\includegraphics[width=0.98\linewidth]{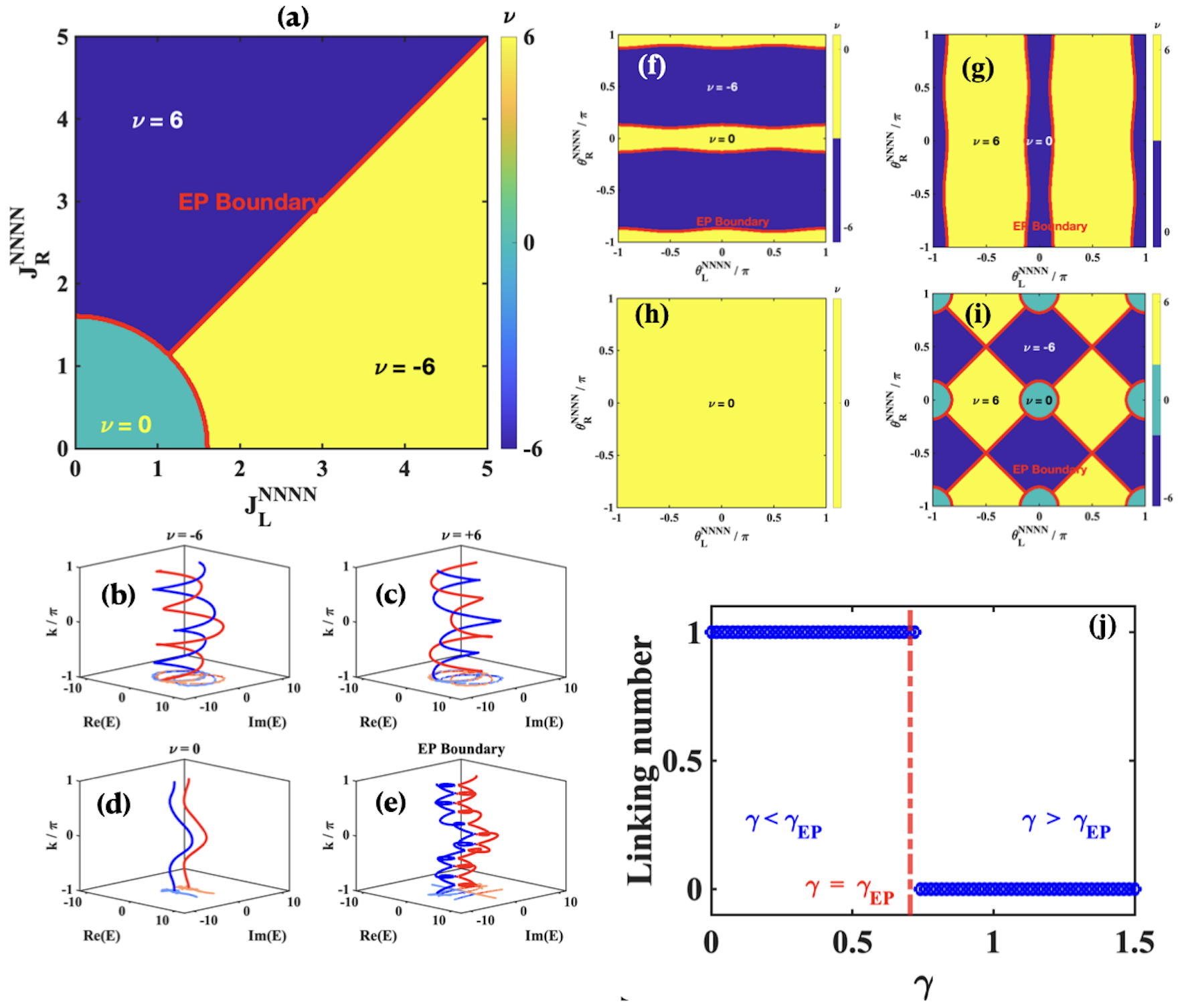}
    \caption{Braiding index in the $J^{NNNN}_L-J^{NNNN}_R$ plane (a) and the evolution trajectories of the particle in the complex plane $Re(E)-Im(E)$ (b,c,d,e) where $E=E_{PBC}=E_{\pm}\in \mathbb{C}$ as in Eq. \ref{eq:model_eq_eig_NNNN2} as the quasimomentum $k$ is varied between $-\pi$ and $+\pi$. Here,\, for (b)\,$P_1 =(J_{L}^{NNNN},J_{R}^{NNNN})=(4.0,1.0)$, (c)$P_2 =(J_{L}^{NNNN},J_{R}^{NNNN})=(1.0,4.0)$, (d) $P_3=(J_{L}^{NNNN},J_{R}^{NNNN})=(0.3,0.3)$, (e) $P_4=(J_{L}^{NNNN},J_{R}^{NNNN})=(3.0,3.0)$. The faded red and blue arrows in the $k=-\pi$ plane denote the handedness of the spectral topology of the bands in (b-e). The braiding index in the $\theta^{NNNN}_L-\theta^{NNNN}_R$ plane for the four points ($P_j, \;j=1,2,3,4$) as indicated in Fig. \ref{fig2}(b-e): (f) $(J_{L}^{NNNN},J_{R}^{NNNN})=(4.0,1.0)$, (g) $(J_{L}^{NNNN},J_{R}^{NNNN})=(1.0,4.0)$, (h) $(J_{L}^{NNNN},J_{R}^{NNNN})=(0.3,0.3)$, (i) $(J_{L}^{NNNN},J_{R}^{NNNN})=(3.0,3.0)$. \, Other parameters:\, $J_{L}^{NN}=J_{R}^{NN}=1.0$,\, $J_{L}^{NNN}=J_{R}^{NNN}=0.7$,\, $\theta_{L}^{NNNN}=\theta_{R}^{NNNN}=-1.5$, \, $\theta_{L}^{NN}=\theta_{R}^{NN}=\theta_{L}^{NNN}=\theta_{R}^{NNN}=0.0$, $\gamma=1.6$. (j) The topological linking number $\mathrm{L}$ plotted against $\gamma$. The red dashed-dotted line refers to $\gamma_{EP}=0.7053$ as in Eq. \ref{eq:EP_NNNN} of the system. Other parameters:\, $J_{L}^{NN}=J_{R}^{NN}=1.0,\, J_{L}^{NNN}=J_{R}^{NNN}=0.7,\, J_{L}^{NNNN}=J_{R}^{NNNN}=0.7, \theta_{L}^{NN}=\theta_{R}^{NN}=0.0, \, \theta_{L}^{NNN}=\theta_{R}^{NNN}=0.0, \;\theta_{L}^{NNNN}=\theta_{R}^{NNNN}=-1.5$. The critical value $\gamma=\gamma_{EP}$ marks the EP at which the linking number undergoes a sharp topological phase transition from 1 to 0. The red curves in (a) and (f-i)denote the EP phase boundaries. These lines represent the exceptional boundaries (exceptional contours) where the complex energy gaps close in the bulk spectrum under PBC.}
    \label{fig2}
\end{figure*}

where, $X_1^2=(J_{L}^{NNNN})^2 \;sin^2{\theta_{L}^{NNNN}} \, e^{6ik}$,
$X_2^2=(J_{R}^{NNNN})^2 \;sin^2{\theta_{R}^{NNNN}} \;e^{-6ik}.$ The Hamiltonian in Eq. \ref{eq:model_eq_eff_NNNN} is $\mathcal{PT}$ symmetric, respecting $\mathcal{PT}H^{NNNN}_{eff}(k)(\mathcal{PT})^{-1}=H^{NNNN}_{eff}(k)$ under reciprocal limits of the NN and NNN couplings $J_L^{NN}=J_{R}^{NN},\theta_L^{NN}=\theta_{R}^{NN}, J_L^{NNN}=J_{R}^{NNN}, \theta_L^{NNN}=\theta_{R}^{NNN}=0,\; \theta_L^{NNNN}=\theta_{R}^{NNNN}$. It is straightforward to show that in the absence of gain/loss non-Hermiticity, EP occurs at an increasingly more number of momenta points at $k_{EP}=({\pm\frac{\pi}{12},\pm\frac{3\pi}{12},\pm\frac{5\pi}{12},\pm\frac{7\pi}{12}}, \pm\frac{9\pi}{12}, \pm\frac{11\pi}{12})$ for $k\in(-\pi,\pi)$ which is a direct consequence of the NNNN interactions via hopping amplitudes and non-Abelian SU(2) phases. Here, the condition for EP to occur, \textit{i.e.} the EP phase boundary is $(J_{L}^{NNNN})^2 \;sin^2{\theta_{L}^{NNNN}} = (J_{R}^{NNNN})^2 \;sin^2{\theta_{R}^{NNNN}}$. The closed-form expression of EP entirely dictated by the NNN hoppings and non-Abelian SU(2) gauge phases is obtained as: 
\begin{equation}
\gamma_{EP}^{NNNN}=\sqrt{\; 2 \;J_{L}^{NNNN} \; J_{R}^{NNNN}\; sin{\theta_{L}^{NNNN}} \; sin{\theta_{R}^{NNNN}}}.
\label{eq:EP_NNNN}
\end{equation}

\begin{figure*}
    \centering
    \includegraphics[width=0.96\linewidth]{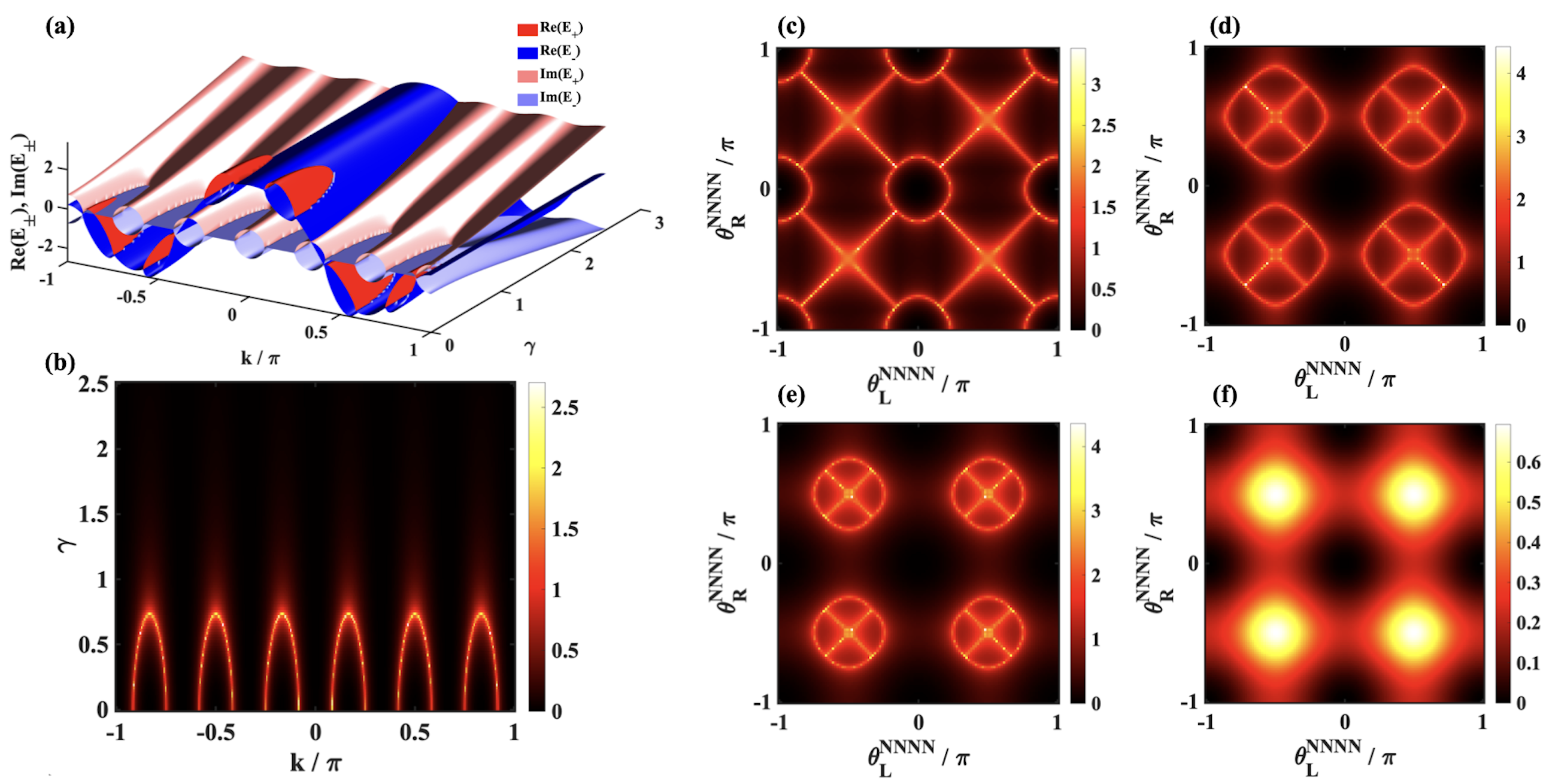}
    \caption{Riemann sheet topology of the eigenvalues and exceptional phase boundaries: (a) $Re(E_\pm)$ and $Im(E_\pm)$ in the $k/\pi-\gamma$ plane; (b) $log_{10} K_{n}$ plotted in the $k-\gamma$ plane; (c-f) $log_{10} K_{n}$ in the $(\theta_{L}^{NNNN}-\theta_{R}^{NNNN})$ plane by varying $\gamma$: (c) $\gamma=0.4$, (d) $\gamma=0.6$, (e) $\gamma=\gamma_{EP}=0.7325$, (f) $\gamma=0.95$. Other parameters: $J_{L}^{NN}=J_{R}^{NN}=1.0$, $J_{L}^{NNN}=J_{R}^{NNN}=0.8$, $J_{L}^{NNNN}=J_{R}^{NNNN}=0.6$, $\theta_{L}^{NN}=\theta_{R}^{NN}=\theta_{L}^{NNN}=\theta_{R}^{NNN}=0.0$. For (a) and (b) $\theta_{L}^{NNNN}=J_{R}^{NNNN}=-2.1$.}
    \label{fig4}
\end{figure*}
\emph{Hopf-link braids and topological phase transition.}--- The spectral braiding topology of the bands of an effective Hamiltonian $H$ is captured by the braiding degree $\nu$ that quantifies how many times the two bands braid around each other in the $(k, Re(E), Im(E))$ space as $k$ varies from $-\pi$ to $+\pi$ \cite{chen2026implementing}:
\begin{equation}
\nu=\frac{1}{2\pi i}\int_{-\pi}^{\pi}\frac{d}{dk}ln \;det (H-\frac{1}{2}Tr(H)).
\label{eq:braid index}
\end{equation}
As shown in Fig. \ref{fig2}(a), in the case of NNNN interactions the braiding indices are found to be $\nu=0,\pm6$. Here, for $\gamma\neq0$ leads to the formation of the $\nu=0$ regime along with the regimes of $\nu=+6$ and $\nu=-6$. If $\gamma=0$, the $\nu=0$ regimes disappear. Moreover, the regimes with $\nu=+6$ and $\nu=-6$ are associated with opposite handedness of the spectral topology of the non-Abelian Hopf-link braids as reflected in Figs. \ref{fig2}(b,c) respectively. On the other hand, for $\nu=0$, the Hopf-link braids vanish and the bands are Abelized, as shown in Fig. \ref{fig2}(d). Along the EP phase boundaries, the bands form distinct nested Hopf-link structures of two interlacing spectral loops, as shown in Figs. \ref{fig2}(e). It undergoes a non-Hermitian topological phase transition across an EP that can be captured by a topological linking number (see Supplementary Information I). Therefore, the formation of Hopf-link braids is a purely exceptional topological signature. 

In addition, Figs. \ref{fig2}(f-i) show the distribution patterns of the braiding indices in the plane of non-Abelian gauge phases $\theta^{NNNN}_L-\theta^{NNNN}_R$ corresponding to the four regimes shown in Fig. \ref{fig2} (b-e). Here, Fig. \ref{fig4}(f,g) show opposite handedness of the spectral topology of the Hopf-braids, as expected. Moreover, Fig. \ref{fig2}(h) shows a flat distribution $\nu=0$. For the EP phase boundary, we have repeating phase patterns of the circular $\nu=0$ and square-like $\nu=\pm6$ regimes. We may also note that the $\nu=0$ regimes appearing in Fig. \ref{fig2}(a,i) disappear as $\gamma\to0$. Including nonreciprocity in the hopping amplitudes via NN, NNN, or NNNN interactions induces an asymmetric phase portrait in the plane of non-Abelian gauge phases $\theta^{NNNN}_L-\theta^{NNNN}_R$. Rich braiding textures are associated with different symmetries (see Supplementary Information III). In Fig. \ref{fig2}(j), the topological linking number is depicted undergoing a sharp phase transition at EP.

\emph{Petermann factor and Eigenvector Coalescence in the Plane of the Non-Abelian Gauge Phases.}--- The Petermann factor quantifies the degree of \textit{non-orthogonality} of the eigenstates of Hamiltonian of a non-Hermitian system \cite{siegman1989excess, wiersig2023petermann, kullig2025generalized}. The phase-space evolution of the braiding index is captured by plotting the logarithmic maximum Petermann factor, $\log_{10}(K_{\max})$, directly in the plane of non-Abelian phases parameterized by $\theta_{\mathrm{L}}^{\mathrm{NNNN}}$ and $\theta_{\mathrm{R}}^{\mathrm{NNNN}}$. Under Periodic Boundary Conditions (PBC), the Petermann factor serves as an eigenvector-centric probe that refers to the coalescence of the eigenstates (with a diverging $K_{\max}$) where the underlying vector space collapses. 

A biorthogonal basis can be used when dealing with a non-Hermitian Hamiltonian ($H \neq H^\dagger$), which often arises in open quantum systems, dissipative systems, or $\mathcal{PT}$-symmetric quantum mechanics. Since $H=H_{eff}^{NNNN}(k)$ is non-Hermitian, its right eigenvectors do not form a usual orthonormal basis. Instead, we use both the right eigenvectors of $H$ and the left eigenvectors of $H$ (which are the right eigenvectors of the adjoint operator $H^\dagger$) to form a biorthogonal basis.

For the $n$-th energy level $E_n$, we have two distinct sets of eigenstates: i) \textit{Right Eigenvectors ($|R_n\rangle$)} which are the standard eigenvectors of the Hamiltonian $H |R_n\rangle = E_n |R_n\rangle$, ii) \textit{Left Eigenvectors ($\langle L_n|$)} which correspond to the conjugate eigenstates, satisfying the eigenvalue equation for the adjoint operator $H^\dagger$, $H^\dagger |L_n\rangle = E_n^* |L_n\rangle$ or  $\langle L_n| H = E_n \langle L_n|$ \textit{($E_n^*$ is the complex conjugate of $E_n$.)}

The right $\{|R_n\rangle\}$ and left eigenstates $\{\langle L_m|\}$ are not orthogonal to themselves, but are mutually orthogonal at different energy levels satisfying the biorthogonality relation: $\langle L_m| R_n\rangle = \delta_{m,n}$,
where $\delta_{m,n}$ is the Kronecker delta ($\delta_{mn} = 1$ if $m=n$, and $0$ if $m \neq n$).
forming a complete biorthogonal basis with the completeness relation expressed as: $\sum_n |R_n\rangle \langle L_n| = I$.

In our system, the Petermann factor can be written as follows in the the n-th left ($\langle L_n|$) and right ($|R_n\rangle$) eigenstates of the effective non-Hermitian Hamiltonian forming a biorthogonal basis $\langle R_m|L_n\rangle=\delta_{m,n}$\cite{wiersig2023petermann, kullig2025generalized}:
\begin{equation}
K_n = \frac{\langle R_n | R_n \rangle \langle L_n | L_n \rangle}{|\langle L_n | R_n \rangle|^2}
\end{equation}
In Fig. \ref{fig4}(a), we have shown the Riemann sheets of the complex eigenvalues $E_{\pm}$ plotted in the $k-\gamma$ plane that connect at the EPs. Fig. \ref{fig4}(b), shows the $K_{\max}=max(K_{n})$ in the $k-\gamma$ plane showing the exceptional trajectories corresponding to $k_{EP}=({\pm\frac{\pi}{12},\pm\frac{3\pi}{12},\pm\frac{5\pi}{12},\pm\frac{7\pi}{12}}, \pm\frac{9\pi}{12}, \pm\frac{11\pi}{12})$ for $k\in(-\pi,\pi)$. The nearest pair of each of these EP phase boundaries later intersects at the exceptional point which in this case turns out to be $\gamma^{NNNN}_{EP}=0.7325.$ In Fig. \ref{fig4}(c-f), four sub-cases are considered corresponding to the different values of $\gamma$ with respect to $\gamma^{NNNN}_{EP}$.
As the staggered imaginary gain-loss non-Hermiticity $\gamma$ is turned on, the system undergoes severe structural transformations. At a low gain-loss non-Hermitian drive ($\gamma = 0.4$), the phase space exhibits an interconnected grid-like network of sharp, highly divergent exceptional lines and circles. This dense network is a direct consequence of multi-harmonic wave interference driven by the competing next-nearest-neighbor (NNN) and third-nearest-neighbor (NNNN) non-Abelian hopping ranges, which split the entire braiding landscape into numerous small, discrete braiding index ($\nu$) sectors. 

As the non-Hermiticity is increased to $\gamma = 0.6$ and $\gamma=\gamma_{EP}= 0.7325$, these exceptional lines undergo geometric reconnections, contracting into four isolated, square-like regimes before constricting tightly into localized circular rings centered around point singularities. Crucially, when $\gamma$ is further increased, all sharp divergent EP phase boundaries completely disappear from the real phase plane, leaving behind only broad, smooth, and low-intensity regimes with minimal overlap values ($K_{\max} \approx 1$). This specific transition signifies that the non-Hermitian gain/loss has a dominating effect on the nonreciprocal hopping amplitudes, forcing the EPs to escape from the real momentum line into the complex momentum plane. Because the Bloch momentum $k$ under PBC is strictly real, it can no longer intersect these complex singularities, causing the energy bands to un-braid into a topologically trivial, completely gapped phase.

It could be worthwhile to mention that the sharp red curves demarcating the edges of the distinct topological sectors represent the exceptional contours of the bulk complex energy spectrum.  Along these trajectories in the $\theta_L^{\text{NNNN}}-\; \theta_R^{\text{NNNN}}$ parameter plane, at least two complex energy eigenvalues become completely degenerate, and their corresponding eigenvectors coalesce, signaling a total breakdown of the diagonalizability of the Hamiltonian. The non-Abelian braiding index $\nu$ can only change its discrete integer value when a gap collapses. Therefore, these red curves mark the loci of EP-mediated topological phase transitions, where the underlying Hopf-link braid undergoes reconfiguration, separating topologically nontrivial regions from the topologically trivial sectors.

\emph{Conclusion and discussion.}--- A novel exotic class of synthetic gauge photonic lattices based on the paradigmatic Hatano-Nelson model is studied with simultaneous considerations of NN, NNN, and NNNN hopping interactions, non-Abelian phases, and gain/loss non-Hermiticity. Multiple competing sources of non-Hermiticity induce novel physical effects, such as higher-order Hopf-link braiding of $\nu=\pm6$, EP-mediated non-Hermitian topological phase transition of the nested Hopf-link structure, and the evolution of exceptional phase boundaries quantified by the Petermann factor. Increasing $\gamma$ beyond EP is shown to lead to the Abelization of non-Hermitian Hopf-link braids and the phase-space structural evolution of $max(K_n)$. 

The Hopf-link braid depicting an interlacing necklace-like spectral structure is fundamentally distinct in origin compared to traditional band braiding, thus generalizing its physical origin and stability \cite{ricca2011gauss, gupta2025non, lapierre2021n, chen2026implementing}. Moreover, similar to applications in anyonic braiding and topological quantum computing \cite{nayak2008non}, Hopf-link braiding with nontrivial linking and braiding may be useful in creating non-Abelian holonomies and linked energy bands \cite{neef2023three} in integrated photonic chips and laser arrays \cite{zhang2022non, mao2026laser} where robust information can be encoded.

The Hopf-link braiding can be achieved in existing experimental settings, including photonic waveguide arrays, microresonators, and electric circuits  \cite{chen2026implementing, gupta2025non}. This could be particularly the case since all the physical elements of the system, namely, the imbalanced hoppings, non-Abelian phases, and onsite gain-loss potentials can be achieved in many experimental platforms. In addition, this is a simplified model with analytically tractable closed-form EP despite the apparent complexity. 

\emph{Acknowledgements.}--- S.K.G. thanks the Department of Physical Sciences, IISER Berhampur for support through the Postdoctoral Research Fellowship.

\emph{Data availability.}--- The data are not publicly available. The data are available from the authors upon reasonable request.

\bibliography{NH_Braids_bib}
\end{document}